\documentstyle[12pt,epsf,epsfig,here]{article}
\def\beq{\begin{equation}}
\def\eeq{\end{equation}}
\def\bea{\begin{eqnarray}}
\def\eea{\end{eqnarray}}
\def\bq{\begin{quote}}
\def\eq{\end{quote}}

\def\NC{{\it Nuovo Cimento} }
\def\NP{{\it Nucl.Phys.} }
\def\PL{{\it Phys.Lett.} }
\def\PR{{\it Phys.Rev.} }
\def\PRL{{\it Phys.Rev.Lett.} }

\def\PTP{{\it Progr.Theor.Phys.} }

\def\gappeq{\mathrel{\rlap {\raise.5ex\hbox{$>$}}
{\lower.5ex\hbox{$\sim$}}}}

\def\lappeq{\mathrel{\rlap{\raise.5ex\hbox{$<$}}
{\lower.5ex\hbox{$\sim$}}}}

\begin{document}
\pagestyle{empty}
\begin{flushright}
{CERN-TH.95-28}
\end{flushright}
\vspace*{5mm}
\begin{center}
{\bf WHERE ARE WE COMING FROM? WHAT ARE WE?} \\
{\bf WHERE ARE WE GOING?} \\
\vspace*{1cm}
{\bf John Ellis} \\
\vspace{0.3cm}
Theoretical Physics Division, CERN \\
CH - 1211 Geneva 23 \\
\vspace*{2cm}
{\bf ABSTRACT} \\ \end{center}
\vspace*{5mm}
\noindent
The Standard Model is the answer to questions 1 and 2, as
established by LEP. Supersymmetry is doubtless the answer to question 3,
as
may well be established by the LHC.
\vspace*{5cm}
\begin{center}
{\it Invited talk at the}\\
{\it Beyond the Standard Model IV Conference}\\
{\it Granlibakken, California}\\
{\it December 1994}
\end{center}

\vspace*{2cm}

\begin{flushleft} CERN-TH.95-28\\
February 1995
\end{flushleft}
\vfill\eject

\setcounter{page}{1}
\pagestyle{plain}

\section{Answers 1 and 2}

The title of this talk is taken from a painting by Paul Gauguin, a
reproduction of which was on my office wall for many years. This meeting
and most
of this talk are devoted to answering the third of Gauguin's questions.
The
answer to his first two questions, namely the Standard Model, has been
verified
by many accelerator experiments, culminating in those at LEP. This
provides the
bedrock which serves as the foundation for our theoretical attempts to
answer
the third question, and may be giving us some hints on the correct
answer, as I
discuss later.

The precision of the LEP results, and these hints, require mastering
some very
subtle experimental effects. By now, it is relatively well-known that
the LEP
determination of the $Z$ mass and width depend on the beam energy, which
is
calibrated using resonant beam depolarization, and is found to be
sensitive to
the phase of the Moon. The tides it induces cause LEP's bedrock to
expand and
contract, affecting the machine's size and hence its beam energy
\cite{aaa}. So
gravity has an effect on LEP - evidence for the unity of physics, if not
for the
unification of fundamental forces! Less well known, perhaps, is the
recent
discovery \cite{bb} that the LEP beam energy is sensitive to how much it
has been
raining, as seen in Fig. 1. More water swells the rock and
expands the machine.
As also seen in Fig. 1, this is also sensitive \cite{bb} to the water
level in
Lake Geneva, albeit with some time delay, much as parts of Northern
Europe are
still rising after the last Ice Age. These effects account for most of
the
variations found previously in the LEP beam energy calibration, and
further improvements in the precision of the data may be possible now
that these
are understood.

\section{Question 3}

Among the questions left unanswered by the Standard Model are the
following.
What is the origin of particle masses? Are they due to the Higgs
mechanism, as
expected by theorists? If so, is the Higgs field composite, as in
technicolour
models? Or is it elementary, in which case is the hierarchy $m_f, M_W,
M_H \ll
m_P$ protected by supersymmetry? Why are there only three fermion
generations,
as we have been assured by LEP? What is the origin of the weak mixing
angles
and CP violation? Are quarks and leptons composite? Or must these
questions
await answers at the string level? Are the strong and electroweak
interactions
combined in a Grand Unified Theory below the Planck scale? If so, are
neutrino
masses and proton decay observable? Are the other particle interactions
unified with gravity in some string theory? Does the quantization of
gravity
entail a modification of the conventional formulations of quantum field
theory
and quantum mechanics?

Presumably Gauguin's third question includes all these, and more. I and
many
other speakers at this meeting would answer these questions within the
framework illustrated in Fig. 2. In the following sections, I will
discuss the
hints from LEP and elsewhere that motivate this framework, and remind
you how
the LHC, in particular, can help answer Gauguin's third question.

\section{Experimental Hints for Supersymmetry}

There are two tentative indications from precision electroweak data,
mainly
from LEP, that favour the supersymmetric worldview. One is the fact that
global fits to the electroweak data tend to favour
\cite{ccc}$^,$\cite{dd} a
relatively light value for the Higgs boson mass: $M_W \lappeq$ 300 GeV
as seen
in Fig. 3. This is consistent with the range predicted
for the mass of the
lightest CP-even Higgs boson in the minimal supersymmetric extension of
the
Standard Model (MSSM): $M_h \simeq M_Z \pm$ 40 GeV \cite{ee}. Moreover,
models
of strongly-interacting Higgs sectors, such as calculable technicolour
models,
are disfavoured by the LEP data. One version of this familiar statement
is shown
in Fig. 4, where a minimal one-generation technicolour model with
$N_{T_c}$ = 2
technicolours and a Majorana technineutrino is confronted \cite{ff} with
the
values of the one-loop radiative-correction parameters
$\epsilon_{1,2,3,b}$
\cite{ggg} extracted from experiment. An attempt to quantify this
discrepancy is
made in Fig. 5, where contours of $\sigma = \sqrt{\Delta \chi^2}$
(corresponding
roughly to the number of standard deviations) are plotted for
one-generation
$N_{T_c} = 2$ models with either a Dirac or Majorana technineutrino: we
see that
$\sigma \gappeq 5$ in both cases.

There has been some discussion at this meeting of the new estimates
\cite{hh}
of $\alpha_{em}(M_Z)$, some of which differ significantly from the
previous best
estimates used in the above analyses of $M_H$ and technicolour models.
We have
made an exploratory study of the possible implications of this increase
in
$\alpha_{em}(M_Z)^{-1}$. In general, an increase in $1/\alpha$
corresponds to a
decrease in $\sin^2\theta_{eff}$, other thing being equal. In fact, LEP
and
other experiments essentially fix $\sin^2\theta_{eff}$, so this effect
must be compensated by a decrease in $m_t$  and/or an increase in $M_H$
and/or a
decrease in $\alpha_s(M_Z)$. We have found \cite{jj} that in an
$\epsilon_{1,2,3,b}$ analysis a fit using just the LEP data, $M_W$ and
the old
$\alpha_{em}(M_Z)^{-1} = 128.87(12)$ is indistinguishable from a fit
including
also the SLC $A_{LR}$ measurement and $\alpha_{em}(M_Z)^{-1} =
129.08(10)$. As
can be seen in Fig. 3, including $A_{LR}$ increases $M_H$ somewhat, but
small
values are still preferred.

The second LEP hint for supersymmetry is provided by the well-publicized
consistency of $\sin^2\theta_W$ with minimal supersymmetric GUTs
\cite{kk}. It is
true that the minimal non-supersymmetric GUT prediction \cite{lll}
\beq
\sin^2\theta_W(M_Z)~_{\overline{MS}} = 0.208 + 0.006 \ln \left({400~{\rm
MeV}\over \Lambda_{\overline{MS}}~(N_f = 4)}\right) = 0.214 \pm 0.004
\label{1}
\eeq
can be excluded. However, the supersymmetric GUT prediction is less
precise,
since it has more parameters. As I discuss later, this means one cannot
use the
value of $\sin^2\theta_W$ to constrain significantly the masses of
supersymmetric particles.

\section{Experimental Constraints on the MSSM}

Let us now discuss the present direct and indirect constraints on the
parameters of the MSSM. Indirect constraints come from the precision
electroweak data discussed earlier, now reanalyzed using MSSM quantum
corrections \cite{ccc}. Figure 6 shows that fits in the MSSM for a given
value of
$M_h$ have values of $\chi^2$ very similar to those in the Standard
Model for the
same value of $M_H$. However, one essential difference is that only a
restricted range of $M_h$ is allowed in the MSSM. Figure 7 shows
$\Delta\chi^2
= 1$ contours for fits to various different selections of electroweak
data, as
well as the bounds on $M_h$ for two values of ${\rm tan}\beta$, the
ratio of
MSSM Higgs v.e.v.'s. The $\Delta\chi^2 = 1$ curves are themselves almost
independent of  ${\rm tan}\beta$ within the corresponding physical
regions.
Notice that in these fits large values have been assumed for $\mu, m_0$
(supposed to be universal) and $m_{\tilde g}$, so that sparticles
essentially
decouple.

We have also explored \cite{ccc} the indirect electroweak constraints on
$m_0$
and $m_{\tilde g}$ (or equivalently $m_{1/2}$), as seen in Fig. 8. These
are
compared with the direct LEP and CDF search limits \cite{mm} for the
same choice
of $\mu, m_A$ and  ${\rm tan}\beta$. We see that the indirect
constraints may be
competitive in some regions of the $(m_0, m_{\tilde g})$ plane. LEP 2,
due to
operate in the years 1996 to 1999, should essentially double the present
direct
LEP lower limit on $m_{\tilde g}$, and the CDF direct lower limit on
$m_{\tilde g}$ should increase to between 300 to 350 GeV within the next
few
years. The CDF limits in Fig. 8 are from a missing energy search: in the
future,
useful limits may also be obtained from searches for the decays of
electroeakly-interacting sparticles into trilepton final states
\cite{nn}.

\section{The Importance of the LHC}

It is clear that the full MSSM parameter space cannot be explored before
the
advent of the LHC. The decision to approve the LHC was taken during this
meeting, and it is clear that my answers to Gauguin's questions would
have been
much less optimistic if it had not been approved. The approval is for an
initial energy of 10 TeV, but I assume that sufficient non-Member-State
support
will become available for the machine to start at the design energy of
14 TeV.
CERN's planning foresees at least four experiments in the initial LHC
programme: two $pp$ discovery physics experiments ATLAS \cite{nn} and
CMS
\cite{oo}, an experiment dedicated to CP violation in $B$ decays
\cite{pp}, and
an experiment ALICE \cite{qq} to look for quark-gluon plasma formation
in
heavy-ion collisions. There may in addition be an experiment to look for
diffractive scattering \cite{rr}, and ideas \cite{ss} for neutrino
experiments
are under active discussion.

Figure 9 demonstrates that the ATLAS experiment \cite{nn} should be able
to
detect strongly-interacting sparticle pair production in the missing
energy
channel for \break
$m_{\tilde g}~\lappeq$~1500~GeV.
The total Standard Model background is not a problem for missing
transverse
energies above about 500 GeV, and the instrumental background is also
expected
to be negligible in this range. Similar sensitivity is to be expected in
the
CMS experiment \cite{oo}. Thus essentially all the parameter space of
the MSSM
allowed by naturalness arguments will be covered. If the LHC does not
discover
supersymmetry, we theorists will have to eat our collective hat.

The prospects of finding the MSSM Higgs sector at the LHC are less clear
\cite{tt}. As is well known and quite visible in Fig. 10a, which shows
the
regions of the $(m_A,$ tan$\beta$) plane accessible to the CMS
experiment
\cite{oo}, there is a troublesome region
100~GeV $\lappeq m_A \lappeq$ 250~GeV, 2 $\lappeq$
tan$\beta \lappeq$ 10 where it will be difficult to discover any of the
MSSM
Higgs bosons. As is seen in Fig. 10b, the ATLAS collaboration \cite{nn}
reckons
that it may ultimately be sensitive in all of the $(m_A,$ tan$\beta$)
plane.
However, there is little safety margin, and this problem requires more
study.

Before leaving the LHC, in view of the interest at this meeting in $B$
physics, it is worth mentioning the physics reach of the LHC for CP
violation
in $B$ decays \cite{pp}. Figure 11 shows as a solid line the present-day
constraints on the CP-violating
 observables $\sin 2\beta$ and $\sin 2\alpha$
inferred from our  knowledge of the CKM matrix \cite{uu}, and the dashed
line
indicates how the constraints may improve by the year 2000. Also shown
are the
likely errors in an $e^+e^-$ $B$-meson factory experiment \cite{vv}, and
what
could be attainable at the LHC \cite{pp}. With error bars as small as
these, in
the next decade flavour physics may become as powerful in testing the
Standard
Model and constraining its possible extensions as are precision
electroweak data
today.

\section{Supersymmetry and GUTs}

The success \cite{kk} of the supersymmetric GUT prediction for
$\sin^2\theta_W$
has already been mentioned. Now I would like to address the question
whether this
success constrains usefully the supersymmetry breaking parameters of the
MSSM. At
the two-loop level, neglecting the uncertainty due to GUT threshold
effects and
retaining just the light thresholds, one has \cite{ww}
\bea
\sin^2\theta_W(M_Z)\bigg\vert~_{\overline{MS}} &=&  0.2029 + {7
\alpha_{em}\over
15\alpha_3}
+ {\alpha_{em}\over 20\pi} \bigg[ -3 \ln \left({m_t\over M_Z}\right) +
{28\over 3}
\ln \left({m_{\tilde g}\over M_Z}\right)
\nonumber \\
&&- {32\over 3} \ln \left({m_{\tilde W}\over
M_Z}\right) - 4 \ln \left({M_H\over M_Z}\right)   - 4 \ln
\left({\mu\over
M_Z}\right) + {3\over 8}~f \bigg]
\label{2}
\eea
where $f$ depends on ratios of supersymmetry breaking parameters and is
about
0.2~$\pm$~0.2, and hence less important numerically than the other
parameters
in (\ref{2}). The relatively precise supersymmetric GUT prediction for
$\sin^2\theta_W(M_Z)\bigg\vert_{\overline{MS}}$ that is often quoted
makes the
assumption that the
unknown MSSM parameters are equal to $M_Z$, or some similar assumption.
One can
invert (\ref{2}) to obtain an expression for the supersymmetry-breaking
gaugino
mass parameter $m_{1/2}$:
\bea
\ln \left({m_{1/2}\over M_Z}\right) &=& {15\pi\over\alpha_{em}} \left[
0.2029 +
{7\alpha_{em}\over 15\alpha_3} - \sin^2\theta_W
(M_Z)\bigg\vert_{\overline{MS}}\right.
-{9\over 4} \ln ~\left({m_t\over M_Z}\right)
\nonumber \\
&&\left.  - {3\over 4} \ln \left({M_H\over
M_Z}\right) - 3 ~\ln~ \left({\mu\over M_Z}\right) + 8.839 + {3\over 8}~f
\right]
\label{3}
\eea
Uncertainties in the quantities on the right-hand side of (\ref{3}),
particularly but not exclusively $\alpha_3(M_Z)$, prevent
\cite{ww},\cite{yy} one
from quoting a meaningfully narrow range for $m_{1/2}$, even if GUT
threshold
effects can be neglected, which is probably not the case. Nevertheless,
the
qualitative agreement with experiment of the minimal supersymmetric GUT
prediction remains impressive circumstantial evidence for supersymmetric
GUTs.

\section{GUTs, Neutrino Masses and Baryogenesis}

There has been much discussion at this meeting of solar neutrino data
and
their interpretation in terms of neutrino oscillations. I certainly
share the
impression that astrophysics alone cannot accommodate the apparent
deficits
found by all the solar neutrino experiments, and find the MSW
matter-enhanced
neutrino oscillation interpretation \cite{zz} with
\beq
\Delta m^2_\nu \sim 10^{-5}~{\rm eV}^2~,\quad\quad \sin^2 ~~2\theta_\nu
\sim
10^{-2} \label{4}
\eeq
the most natural. Perhaps these are the first direct indications of
physics
beyond the Standard Model?

Theoretically, the most appealing model for neutrino masses is the GUT
see-saw
matrix:
\beq
\left(\matrix{\sim 0 & m_f \cr  m_f & M_M}\right)
\label{5}
\eeq
where $M_M$ is a Majorana mass for the right-handed neutrino. This
suggests that
\beq
m_{\nu_e}~:~m_{\nu_{\mu}}~:~m_{\nu_{\tau}} \sim m^2_u~:~m^2_c~:~m^2_t
\label{6}
\eeq
assuming there is not a large hierarchy in the $M_M$ for different
generations,
in which case the MSW solution (\ref{4}) suggests that
\beq
m_{\nu_e} \ll m_{\nu_{\mu}} \sim (2~{\rm or}~3) \times 10^{-3}~{\rm eV}
\label{7}
\eeq
Scaling this up by $m^2_t/m^2_c$, it appears perfectly reasonable to
expect
that
\beq
m_{\nu_e} \sim 10~{\rm eV}
\label{8}
\eeq
as advocated by enthusiasts for a component of Hot Dark Matter.

If this model is correct, evidence for it may soon be found in
accelerator
experiments. The two CERN experiments (CHORUS and NOMAD) designed to
look for
$\sin^2\theta_{e\mu} \gappeq 10^{-4}$ when $\Delta
m^2_{\nu_\mu,\nu_\tau}
\sim 10^2$ eV$^2$ are now operating, and many GUT see-saw models predict
$\nu_\mu - \nu_\tau$ mixing angles within their range of sensitivity
\cite{aai}.
Will they provide the first laboratory evidence for physics beyond the
Standard
Model?

If the estimate (\ref{8}) is correct, and we use (\ref{5}) with $m_t
\sim$ 170
GeV, we need $M_M \sim 10^{12}$ GeV for the third-generation
right-handed
neutrino mass. This is considerably below the supersymmetric GUT scale
of
10$^{16}$ GeV, but the appearance of right-handed neutrinos in this mass
range
would not upset  the calculation of
$\sin^2\theta_W(M_Z)\bigg\vert_{{\overline{\rm MS}}}$, since they
$SU(3)\times
SU(2)\times U(1)$ gauge singlets. Indeed, such a value of $M_M$ could be
boon
to cosmological baryogenesis.

To my mind, the most elegant scenario \cite{bbi} for this
is $\nu_R\rightarrow L + H$ decay producing a net lepton asymmetry
$\Delta L
\not= 0$, which is then recycled by non-perturbative electroweak effects
with
$\Delta (B-L) = 0$ to yield finally a net baryon asymmetry
$\Delta B \not= 0$. This works only
if the $\nu_R$ are produced after inflation, which requires inflaton
$\Phi
\rightarrow \nu_R$ decay, and hence $M_M < m_\Phi$. The COBE observation
of
fluctuations in the microwave background radiation suggests that $m_\Phi
\sim
10^{13}$ GeV, in which case $M_M \lappeq 10^{12}$ GeV is required, and
the
neutrino masses cannot be much smaller than the
astrophysically-preferred
values (\ref{7}), (\ref{8}).
Thus the MSW
interpretation of the solar neutrino data is compatible not only with
the
$\nu_\tau$ constituting a Hot Dark Matter component, but also with
neutrino
baryogenesis \cite{cci}.

\section{Answer 3: Superstring}

This is the only candidate we have for a Theory of Everything, including
quantum gravity, and hence for our ultimate destination beyond the
Standard
Model. However, as you know, there is considerable ambiguity in the
choice of
string model, and hence a frustrating ambiguity in its experimental
predictions. In the past, people have constructed
string models based on non-unified gauge groups such as $SU(3)^3$
\cite{ddi},
$SU(3)\times SU(2)\times U(1)^n$ \cite{eei} and $SU(5)\times U(1)^m$
\cite{ffi}.
The latter flipped $SU(5)$ model has been my personal interest: it is
the
closest to a conventional GUT that can be constructed without using a
higher-level representation of the Kac-Moody current algebra on the
world sheet
\cite{ggi}. Progress has recenbly been made in formulating higher-level
models
\cite{hhi}. Though a completely realistic example has yet to emerge, it
may be
possible in this way to construct a realistic superstring GUT. I will
not enter
into specific models here, but conclude this talk by reminding you of
two
interesting qualitative predictions of string models that are relatively
model-independent.

	One is the calculation \cite{jji} of the string unification scale,
i.e., the
energy at which the extrapolated low-energy gauge couplings should
appear to
become equal, which is
\beq
M_{SU} \simeq 5 \times g \times 10^{17}~{\rm GeV}~\times F
\label{9}
\eeq
where $g$ is the gauge coupling and $F$ depends on the specific string
model
chosen, which is about unity in models constructed out of free
world-sheet
fermions \cite{kki}. The prediction (\ref{9}) appears to be somewhat
larger than
the minimal supersymmetric GUT calculation of about 10$^{16}$ GeV.
Perhaps we
should look at models with $F < 1$, or perhaps we should look at models
with
additional light particles, or perhaps the GUT unification scale really
is
below below the string unification scale, as occurs in flipped $SU(5)$.
It is
encouraging that string at least gives us a unification scale ot aim at.

The second qualitative string prediction I would like to emphasize is
that for
$m_t$. It is a generic feature of models derived from string that
non-zero
Yukawa couplings $\lambda$ are of the same order as the gauge coupling
\cite{lli}. Specifically, in free-fermion models such as flipped $SU(5)$
\cite{ffi} \beq
 \lambda = \sqrt{2} g
\label{10}
\eeq
If applied to the top quark Yukawa coupling, this yields after
renormalization
a value of $m_t$ below but close to the approximate infrared fixed
point:
\beq
m_t \simeq 190 \sin\beta ~~{\rm GeV}
\label{11}
\eeq
in the case of free fermion models. Not such a bad prediction! There are
many
other interesting ideas circulating about Yukawa
unification \cite{mmi} and the possible dynamical determination of the
top and
other quark masses \cite{nni}, which I do not have time to discuss here.

Physicists sometimes despair of ever being able to prove that string is
the
answer to Gauguin's third question, even if it is. These two examples
may serve
as some encouragement that testing string may not be impossible, even in
the
absence of direct probes of quantum gravity.

\vfill\eject

\newpage
\noindent
{\bf FIGURE CAPTIONS}


\noindent
{\footnotesize Fig. 1 - Sensitivity of the LEP beam energy to (a) tides
[1]: the
solid lines are due to a tidal model,  (b) the water table in the Jura
mountains and (c) the level of Lake Geneva [2].}


\noindent
{\footnotesize Fig. 2 - The supersymmetric worldview answers Gauguin's
third question.}


\noindent
{\footnotesize Fig 3 - Contours of $\Delta\chi^2 = 1$ in the $(M_H,m_t)$
plane for a
Standard Model analysis [3] of all electroweak data (ALL), including (+)
or not (-) the SLD measurement of $A_{LR}$ (ALR) and the CDF kinematic
fit to
$m_t$ (CDF). The $\Delta\chi^2 = 1$ bands allowed by CDF and ALR alone
are shown
separately. For ALL$^{+\rm CDF}_{\rm -ALR}$ fit, which combines the LEP,
CDF
and low energy data, the $\Delta\chi^2 = 4(2\sigma )$ contour is also
shown
.}


\noindent
{\footnotesize Fig. 4 - Comparison [6] of the Born approximation
(stars), projections of
the $\Delta\chi^2 = 1,4$ ellipsoid (solid ellipses), the SM (grid) and
the
predictions of a one-generation TC model with $N_{TC}$ = 2, a Dirac
technineutrino, $M_U = M_D$, 100 GeV $< M_E <$ 600 GeV, 50 GeV $< M_N <
M_E$ (scattered dots). The TC predictions are added to the
SM radiative corrections, using the reference values $m_t$ = 170 GeV and
$M_H
= M_Z$. Note that the TC predictions are further than the SM from the
experimental data. The bold arrows labelled TQ and B indicate possible
shifts
in the TC predictions of definite sign, and the other (thin) arrows
labelled
B and NC indicate shifts that are less certain.}



\noindent
{\footnotesize Fig. 5 - Contours [6] of
$\sigma\equiv\sqrt{\Delta\chi^2}$ for
one-generation models with either Dirac technineutrinos (a), (b) or
Majorana
technineutrino (c), (d). Note that $\sigma \gappeq$ 4.5 in all of the TC
parameter space, to be compared with $\sigma$ = 2.6 in the SM at the
reference
point ($m_t$ = 170 GeV, $M_H = M_Z$). In the case of techniquark mass
degeneracy
($M_U = M_D)$, the Dirac and Majorana models fits are comparable; in the
case
$M_U > M_D$, however, the Dirac model becomes highly disfavoured.}


\noindent
{\footnotesize Fig. 6 - Curves of $\chi^2$ as function of the (lightest)
Higgs boson
mass in the Standard Model (dashed line) and the MSSM (solid line) at
$m_t$ =
150, 170, 190 GeV, using ALL$^{\rm +CDF}_{\rm -ALR}$ data [3]. In the
MSSM case
we choose $\tan\beta = 4, \mu = m_{\tilde g} = m_0$ = 1 TeV. Notice that
the
dashed and solid curves are very close and finally merge when $m_h$
reaches
its theoretical upper limit.}


\noindent
{\footnotesize Fig. 7 - Contours of $\Delta\chi^2$ = 1 in the $(m_h,
m_t)$ plane for an
MSSM analysis [3] of all electroweak data (ALL), including (+) or not
(-)
the SLD measurement of $A_{LR}$ are set to large values by choosing $\mu
m_{\tilde g} = m_0$ = 1 TeV. Also shown are the theoretical lower and
upper
bounds on $m_h$ for $\tan\beta$ = 2 and 16, corresponding to $m_A = 0,
\infty$.
Actually each curve is slightly dependent on $\tan\beta$ within the
allowed
range; the difference is however negligible for our purposes, and a
smoothed
average is shown.}


\noindent
{\footnotesize Fig. 8 - Exclusion plot [3] in the $(m_0, m_{\tilde g})$
plane for $m_t$
= 165 GeV, $\mu$ = -250 GeV, $m_A$ = 500 GeV, $\tan\beta$ = 2. The solid
curve
labelled ``MSSM R.C." encloses the region excluded by our ALL$^{\rm
+CDF}_{\rm
ALR}$ fit (with MSSM radiative corrections) at $\Delta\chi^2$ = 2.7
(90\% C.L. on
each variable separately). Also shown are the limits on slepton and
chargino
masses from LEP, and the exclusion contours from the negative results of
CDF
searches for gluinos and squarks (CDF solid line: with cascade decays;
CDF
dashed line: no cascade decays; the cusp corresponds to the case
$m_{\tilde g} =
m_{\tilde q}$. Notice that the region just above the chargino threshold
$M_Z$/2
is here excluded both by CDF and the MSSM R.C. analysis (assuming the
standard
mass relations in the MSSM).}


\noindent
{\footnotesize Fig. 9 - Missing-energy signature of squarks and gluinos
in the ATLAS
[13] detector (histogram) compared with the physics background (open
circles), na\"\i ve estimate of instrumental background (solid squares)
and
more realistic estimate (open triangles).}



\noindent
{\footnotesize Fig. 10 - Search for MSSM Higgs bosons in the $(m_A,
\tan\beta)$ plane
with (a) CMS [14], (b) ATLAS [13].}

\noindent
{\footnotesize Fig. 11 - Present and possible future constraints within
the Standard
Model on the CP-violating parameters $(\alpha , \beta)$ that are
measurable in
$B$ decays [20], compared with the superweak theory prediction and
possible
future measurement errors.}

\end{document}